\documentclass[fleqn,usenatbib]{mnras}

\usepackage{newtxtext,newtxmath}

\usepackage[T1]{fontenc}

\DeclareRobustCommand{\VAN}[3]{#2}
\let\VANthebibliography\thebibliography
\def\thebibliography{\DeclareRobustCommand{\VAN}[3]{##3}\VANthebibliography}

\usepackage{graphicx}	
\usepackage{amsmath}	
\usepackage{xcolor} 
\usepackage{hyperref}
\usepackage{derivative}

\title[Outcomes of Planetary Collisions]{Outcomes of Planetary Collisions: Importance of Gravity and Material Properties}

\author[Smallwood et al.]{
Jeremy L. Smallwood,$^{1,2}$
Jeffrey S. Lee,$^{1}$
Lorin S. Matthews,$^{1}$
and Bryant M. Wyatt,$^{1,3}$\thanks{E-mail: wyatt@tarleton.edu}
\\
$^{1}$Center for Astrophysics, Space Physics, and Engineering Research, Baylor University, Waco, Texas 76798-7316, USA\\
$^{2}$Institute of Astronomy and Astrophysics, Academia Sinica, Taipei 10617, R.O.C. \\
$^{3}$Department of Mathematics, Tarleton State University, Stephenville, TX USA\\
}

\date{Accepted XXX. Received YYY; in original form ZZZ}

\pubyear{2015}

\begin{document}
\label{firstpage}
\pagerange{\pageref{firstpage}--\pageref{lastpage}}
\maketitle

\begin{abstract}
The final sizes, composition, and angular momenta of solid planetary bodies depend on the outcomes of collisions between planetary embryos.  The most common numerical method for simulating embryo collisions is to combine a gravity solver with a hydrodynamic solver, allowing pressure gradients, shock waves, and gravitational torques to loft material into orbit.  Here, we perform the first direct comparison between hydrodynamic methods and a simplified method employing only gravity and a quadratic repulsive force.  The formation of Earth's Moon, perhaps the most heavily simulated planetary collision, is used as a test case.  Many of the main features of a collision between two planetary embryos, including collisions in which an orbiting disc of material and/or intact moons are formed, are controlled solely by gravitational forces. Comparison of the methods shows that the mass and orbit of the satellite, as well as the extent of physical mixing between the protoearth and impactor, are similar regardless of the inclusion of the inclusion of hydrodynamic effects or the equation of state employed. The study of thermal and chemical effects of the impact, and determining the time scale for lunar accretion, still require a full hydrodynamic calculation. The simplified gravity plus quadratic repulsive force approach allows rapid testing of various initial conditions to identify cases for further detailed study.
\end{abstract}




\section{Introduction}

The late stage of terrestrial planet formation is thought to be dominated by giant impacts which can affect the masses, compositions, locations, and spin rates of terrestrial planets \citep{Morbidelli2012}. Collisions between planet-scale objects have played a role in setting the initial thermal and chemical states for the evolution of each terrestrial planet in the solar system. For example, Mercury may have experienced a collision that removed much of its silicate mantle \citep{Benz2007}, while Venus may have experienced one or more collisions that affected its spin rate \citep{Alemi2006}.  A planetary-scale collision has also been proposed as an explanation for the north/south crustal dichotomy on Mars \citep{Marinova2008}. Finally, the Moon may have formed as a result of a planetary-mass impactor striking the proto-Earth \citep{Hartmann1975,Cameron1976}. Studying the dynamics of giant impacts sheds light on the late stages of planetary accretion and the formation of Moon-like bodies.

Early efforts to identify the type of impact that could have formed Earth's moon used smoothed particle hydrodynamics methods to try to reproduce three observed characteristics of the system: (1) the masses of the Moon and Earth, (2) the angular momentum of the system, and (3) the Moon's low iron content \citep{Benz1986,Benz1987,Benz1989,Cameron1991,Cameron1997}.  As computing power increased over the course of these studies, each successive attempt used more realistic equations of state (relating pressure, density, temperature, and internal energy for each material), and increased numerical resolution.  \citet{Canup2001} identified a class of potential impacts: an oblique collision between a Mars-sized planetary embryo and a body slightly less massive than the Earth at near escape-velocity.  The precise conditions for which the outcome of the collision satisfied observations (1), (2), and (3) were refined slightly over the ensuing decades \citep{Canup2004,Canup2008,Cuk2012,Lock2017}, and additional considerations (e.g., pre-impact rotation of the colliding bodies \citet{}) were also explored.

In recent years, geochemical data obtained from lunar samples show that the isotopic ratios of many species including oxygen \citep{Epstein1970, Wiechert2001}, chromium \citep{Lugmair1998}, tungsten \citep{Touboul2007}, and silicon \citep{Georg2007,Zhang2012} are similar to Earth's mantle.  Three hypotheses have been proposed: the protoearth and impactor had similar isotopic ratios \citep{Wiechert2001, Mastrobuono2015, Kaib2015, Dauphas2017}, the isotopic ratios were initially different and equilibrated in a disc \citep{Pahlevan2007, Zhang2012}, or the objects mixed significantly during the impact event \citep{Cuk2012, Canup2012, Young2016,Nielsen2021}, resulting in an orbiting disc rich in terrestrial material.

In the ``canonical'' Moon-forming impact scenarios developed by \citet{Canup2004}, the majority of material in the impact-generated disc, the material from which the Moon is formed, originates with the impactor.  This has motivated the development of alternative impact scenarios in which a fast, small impactor strikes a rapidly spinning protoearth \citep{Cuk2012}, or two like-sized objects strike each other \citep{Canup2012}.  Both of these scenarios create a disc rich in terrestrial material which could readily explain the isotopic ratios.  However, they create systems with angular momenta initially higher than the present-day Earth/Moon system, and rely upon an overly simplified tidal evolution scenario \citep{Wisdom2015} to shed excess angular momentum.  

Regardless of the impact scenario, there are lingering concerns about the ability of numerical impact simulations to properly represent mixing in giant impacts \citep{Melosh2009,Hosono2016, Deng2019}.  The vast majority of Moon-forming impact simulations have been performed using the smoothed particle hydrodynamics (SPH) technique (see, e.g., \citet{Monaghan1992} for discussion).  This numerical method creates an artificial tension force at boundaries between materials of differing densities (i.e., the core/mantle boundary) which inhibits mixing (see e.g., \citet{Hosono2016} and references therein). The artificial tension force inhibits mixing between core and mantle in Giant Impact simulations, and also affects the efficiency with which debris re-accretes onto the proto-Earth after the impact \citep{Deng2019}.  

Given this limitation of SPH-codes, it is highly desirable to explore alternative numerical techniques and to determine whether the extent of mixing between the bodies and provenance of disc material changes significantly when different numerical techniques are used.  Here we perform simulations of the a standard "benchmark" case for the Moon-forming impact \citep{Barr2016} between iron/silicate impactor and target using a standard SPH method \citep{CanupBarr2013}, an adaptive SPH method \citep{Owen1998}, an Eulerian/Lagrangian hydrocode (CTH) \citep{McGlaun1989,Crawford2006,CanupBarr2013}, and a novel gravitational code that balances gravity and the short-range repulsive force arising from physically overlapping particles \citep{Eiland2014}.  We find that regardless of material equation of state or solution method, all four methods yield similar orbiting disc masses, scaled angular momenta, and as a result, final satellite masses close to the mass of the Moon.  This suggests the mass of the impact-generated disc created in a planetary collision could be constrained by methods which consider only the gravitational forces, similar to those presented in \cite{Eiland2014} and \cite{Wyatt2018}, without the need for a full hydrocode simulation. However, each of the four methods yield different results in terms of the mass of iron in the Moon, and the extent of mixing between impactor/target during the impact.  The importance of this gravity-centric model is rooted in its capacity to illustrate that orbital behaviors, encompassing attributes like spin, angular momentum, and mass, are primarily governed by gravitational forces.  The resulting characteristics of a system are modified by the material properties of the objects, which are modeled by an appropriate equation of state. In the gravity-centric model, the complicated network of equations describing the state changes is replaced by a simplified inelastic spring model.

The layout of the paper is as follows. Section~\ref{sec::methods} provides a description of each of the codes used to model the impact (given a specific set of initial conditions) and the methods used to analyse the results, which are presented in Section~\ref{sec::results}. The implications of these results and potential application for future studies are give in in Section~\ref{sec::summary}.


\begin{table}
\caption{Initial conditions for benchmark giant impact \citet{Canup2004}.}
\centering
\begin{tabular}{l l l}
\hline
 Parameter  & Symbol & Value \\
\hline
Total Mass & $M_{\rm T}$ & $1.025\, \rm M_{\oplus}$ \\
Impactor-to-total Mass Ratio & $\gamma$ & $0.13$ \\
Target Mass & $M_{\rm tar}$ & $0.8875\, \rm M_{\oplus}$\\
Target Radius & $R_{\rm tar}$ & $6.30\times 10^3\, \rm km$\\
Impactor Mass & $M_{\rm imp}$ & $0.1375\, \rm M_{\oplus}$ \\
Impactor Radius & $R_{\rm imp}$ & $3.58\times10^3\, \rm km$ \\
Impact angle & $\theta$ & $50.9^{\circ}$ \\
Impact velocity & $v_{\rm imp}$ & $8.34\, \rm  km/s$ \\
Target $\hat{x}$ velocity & $v_{\rm x,tar}$ & $1.09\, \rm km/s$\\
Impactor $\hat{x}$ velocity &  $v_{\rm x,imp}$ & $7.26\, \rm km/s$ \\
\hline
\label{table:initial_conditions}
\end{tabular}
\end{table}

\section{Methods}
\label{sec::methods}

\subsection{Initial Conditions}
To compare the results of giant impacts across several different numerical methods, we chose a common set of initial conditions corresponding to ``run119'' of \cite{Canup2004}, an oblique collision between a protoearth slightly less massive than the final Earth and an impactor that is roughly the size of Mars. Because this simulation was among the first to create an Earth-Moon system with the proper mass, angular momentum, and lunar iron fraction, run119 has been used as a standard ``benchmark'' case for giant impact simulations \citep{CanupBarr2013,Barr2016,Deng2019}.  Using standard parameters from run119 also allows us to compare results from our current simulations to those published in prior works \citep{Canup2004,CanupBarr2013}.

Table \ref{table:initial_conditions} summarizes the initial conditions used in all of our simulations. The total mass involved in the impact is $M_{\rm T}=1.02M_{\oplus}$ where $M_{\oplus}=5.98 \times 10^{24}\, \rm kg$ .  The mass of the "target" (the protoearth) is $M_{\rm tar}=0.88 M_{\oplus}$, and the mass of the impactor is $M_{\rm imp}=0.13M_{\oplus}$, giving rise to an impactor-to-total mass ratio $\gamma=0.13$.  Each body is composed of 30 per cent iron and 70 per cent silicate by mass, with reference densities $\rho_{\rm Fe}=7800\, \rm kg/m^3$ and $\rho_{\rm Si}=3320\, \rm kg/m^3$.  With these densities, the radius of the target object is typically around $6300\, \rm km$, and radius of the impactor approximately $3580\, \rm km$. However, it is not always possible to achieve identical values of initial mass and radius across the various numerical methods, owing to differences in the equation of state and the manner in which the objects achieve hydrostatic equilibrium before the impact.  The method used to implement these conditions in the SPH, CTH, and {\sc cataclysm} codes are described below.

\subsection{SPH}
Smoothed particle hydrodynamics is the most common approach used to simulate the Moon-forming impact \cite[e.g.,][]{Benz1986,Canup2001,Canup2004}.  Technical details regarding standard SPH and its application to the impact can be found in, e.g., \citet{Canup2004} so we will only provide an overview here.  In SPH, material in the domain is broken into equal-mass particles, whose radii are calculated based on their thermodynamic states, i.e., their densities.  The evolution of the group of particles is calculated based on the balance between pressure forces, gravitational forces, and shock dissipation.  The governing equations for SPH describe the motion of a fluid and require an equation of state, which describes the pressure of a material (or mixture of materials) as a function of density and internal energy.  

In addition to the standard SPH formulation, we have also performed simulations using a new version of SPH, Adaptive SPH, in which the kernel functions describing the SPH particles can be elliptical, allowing better representation of highly anisotropic flows \citep{Owen1998}.  This method has been implemented in the open-source hydrocode Spheral++\footnote{http://sourceforge.net/projects/spheral/}.

 In both the standard SPH formulation and Spheral++, one specifies a mass and radius for the core and mantle for the initial target and impactor, but each object must be ``settled,'' by allowing it to remain motionless in an empty domain so that a self-consistent hydrostatic state can be determined based on the equation of state and the initial temperature of the object (assumed in this case to be uniform).  The object is settled until the kinetic energy and angular momentum remain roughly constant between time steps, which can take perhaps 3 to 4 hours of real time.  

Two equations of state are commonly used in Moon-forming impacts: the Tillotson equation of state \citep{Tillotson1962} and ANEOS \citep{Thompson1972}.  Both equations of state are analytic.  Tillotson is designed to mimic the linear shock velocity-particle velocity equation of state at low pressures and the behavior of a Thomas-Fermi model at high pressures \citep{Melosh1989}.  The major drawback of using the Tillotson equation of state to simulate the Moon-forming impact simulations is that it cannot provide phase information, and can occasionally return non-physical results at low temperatures and high pressures, which can arise, for example, in the iron core of the protoearth if the initial temperature is below a few thousand Kelvin.

ANEOS is based on calculation of the Helmholtz free energy of the materials, which can be broken into three separate parts arising from non-temperature-dependent interactions, temperature-dependent interatomic forces, and energy related to the ionization of atoms \citep{Thompson1972}.  A major shortcoming of ANEOS was its assumption that vapor is composed purely of atoms, rather than molecular clusters, but this has been corrected by \cite{Melosh2007}, and the version of ANEOS including molecular clusters (MANEOS) is now commonly used in Moon-forming impact calculations with SPH (e.g., \citealt{Canup2004}) and CTH \citep{CanupBarr2013}.  

To compare results from ``standard'' SPH to those from other methods, we use information about the outcomes of run119 performed by \cite{CanupBarr2013}, which employed the MANEOS equation of state with numerical resolutions of $N = 10^4$ (SPH-M(LR)) and $N = 10^5$ (SPH-M(HR)) particles.  Details about the equation of state and simulation methods may be found in their paper.  We have also performed run119 using Spheral++ and a Tillotson equation of state (SPH-T).  We use the Tillotson equation of state for basalt and iron to represent the silicate mantles and iron cores of the protoearth in our simulations.  

\subsection{CTH}
The CTH shock physics code has been used extensively by the planetary community to simulate the formation of impact craters on planetary surfaces \cite[e.g.,][]{McGlaun1989,Pierazzo1997,Pierazzo2000,Pierazzo2008,Syal2015}.  CTH solves the governing equations of shock propagation and material deformation using an iterative approach in which material flow is modeled using a deformable mesh, then re-cast as a flow through a static Eulerian mesh to provide the physical and thermodynamic state of each parcel in the domain.  CTH has been modified to include self-gravity and adaptive meshing \citep{Crawford2006}, and was shown to produce outcomes for the Moon-forming impact similar to those achieved with conventional SPH methods, including for run119, the test case used here \citep{CanupBarr2013}. 

 In CTH, one specifies the target and impactor radius and the radii of their cores, but not the total mass of the objects.  Before the impact occurs, a hydrostatic equilibrium must be achieved in both bodies, and so the masses of the objects are adjusted until a self-consistent and stable pressure and temperature profile can be achieved.  This may require changing the masses of the bodies by a few percent.  The pressure, temperature, and density profiles depend on the equation of state, and so the masses of the objects may change by a few percent depending on the initial temperature profile specified. Special care was taken to use impactor and target masses, iron fractions, and iron/silicate equations of state identical to  those in run119  from \cite{Canup2004}.  Details about the methods used to simulate the Moon-forming impact in CTH, including the values of MANEOS parameters, the initial conditions, and outcomes, can be found in \cite{CanupBarr2013} and \cite{Barr2016}. The simulations were performed in a 3-dimensional Cartesian domain 125$R_{\oplus}$ on a side, to ensure that the overwhelming majority of mass remains inside the domain.  Adaptive meshing permits smaller elements to be used in regions of high density, and element sizes were assigned based on local density, so that each element has roughly equal mass.  This mimics the equal-mass particles used in SPH simulations.

\subsection{{\sc cataclysm}}
Richard Feynman once asked, "If, in some cataclysm, all of scientific knowledge were to be destroyed, and only one sentence passed on to the next generation of creatures, what statement would contain the most information in the fewest words?" He answered by stating, "I believe it is the atomic hypothesis (or the atomic fact, or whatever you wish to call it) that all things are made of atoms — little particles that move around in perpetual motion, attracting each other when they are a little distance apart, but repelling upon being squeezed into one another. In that one sentence, you will see, there is an enormous amount of information about the world, if just a little imagination and thinking are applied" \citep{Feynman1965}.  Inspired by this statement, we created an "atomistic" model of gravitationally interacting bodies which attract each other when separated by distances greater than the particle diameters but repel each other when squeezed together.  The goal here is to produce a model that is computationally inexpensive, yet maintains enough sophistication to capture the major components of planetary impacts. 

The model is designed around three main components: the weak but ever present attractive force of gravity, the strong but short-ranged repulsive force caused by objects in physical contact, and the transformation of energy. The dynamics of the model are solely determined by the sum of all the pairwise forces between elements: (1) elements always exert a gravitational force upon each other, (2) if two elements are not in physical contact, gravity is the sole force, (3) if two elements come into physical contact, they impart an additional strong repulsive force on each other, and (4) this repulsive force is non-elastic to remove energy from the collision due to a large fraction of the collisional energy being converted into other forms. 

The attractive  gravitational force is modeled by treating each spherical element with diameter $D$ as a point mass such that the magnitude of the force between elements with masses $M_{\rm i}$ and $M_{\rm j}$ is given by $F_{\rm ij} = G M_{\rm i} M_{\rm j}/r_{\rm ij}^2$, where $G$ is the gravitational constant and $r_{\rm ij}$ is the distance between the elements' centers.  When two elements come into contact, the elements experience a repulsive force.  The contact region of two overlapping spheres is a circle, with an area determined by the separation between the elements' centers. Thus, this repulsive force is proportional to the square of the separation between elements \citep{Mravlak2008} and is given by
$(K_{\rm i} + K_{\rm j})(D^2 - r_{\rm ij}^2)$, where $K_{\rm i}$ is a parameter which characterizes the strength of the repulsive force  for a given material. While elements in contact are under compression, the repulsive force is large. When this repulsive force overcomes element compression and the elements begin to move apart, the repulsion parameter is reduced by a given percentage using a multiplier $\alpha_{\rm i}$, such that the repulsive force is $(\alpha_{\rm i} K_{\rm i} + \alpha_{\rm j} K_{\rm j})/(D^2 - r_{ij}^2)$. This produces a non-elastic collision, to encapsulate the collision energy that is lost due to its transformation into all other forms. The details of how $K_{\rm i}$ and $\alpha_{\rm i}$ are related to the thermodynamics are given in the Appendix.

In {\sc cataclysm}, the masses of the impactor and target, along with their mass fractions of iron and silicate, are used to determine a common radius for all computational-elements based on the number of computational-elements specified. The impactor and target are created by randomly placing the core-forming iron computational elements into two separate spherical domains surrounded by a mantle of silicate computational elements.  The radii of each region are calculated by doubling the radii determined by a 68\% packing ratio. The objects are settled in a procedure similar to that used in SPH. The elements of each body interact under a large damping constraint for 20 simulated hours, then the damping constraint is removed and the bodies allowed to reach equilibrium over 50 simulated hours. Next the bodies are then spun up based on initial input for spin vectors and angular momenta for an additional 50 simulation hours. Depending on the total number of elements, this entire setup can take from a few seconds to a few hours. For example on a laptop equipped with an NVIDIA GeForce RTX 3080 GPU, the setup for a run with 1024 computational elements takes less than 30 seconds and a setup with 131,072 computational elements takes around 2 hours and 45 minutes. At this point, the angular momentum of the target and impactor are determined and initial positions and velocities for the target and impactor can be selected so that the entire system can have any prescribed angular momentum.

\begin{figure}
\centering
\includegraphics[width=1\columnwidth]{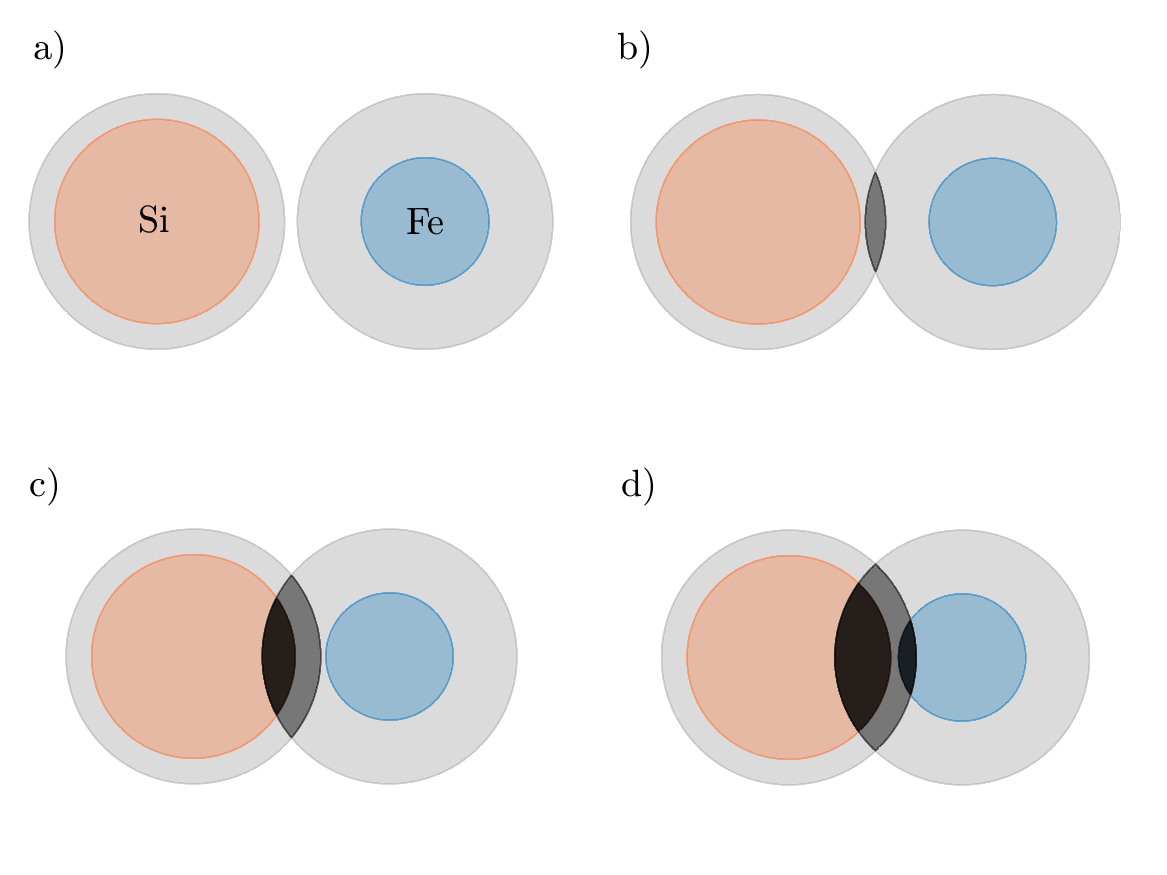}
\caption{Possible interactions between two elements in {\sc   cataclysm}. The silicate (Si) element is red with a gray shell. The iron (Fe) element is blue with a gray shell. a) Elements are not in contact and only gravitational forces are present. b) Elements are in contact but
only the shells overlap; the repulsive force is elastic. c) Silicate shell has been penetrated, but not the iron shell. The repulsive parameter associated with the silicate element is $K_{\rm Si}$ as the particles approach and $\alpha_{\rm Si} K_{\rm Si}$ as the particles separate. d) Both shells have been penetrated and both repulsive parameters are reduced when the particles separate.}
\label{fig:two_ele}
\end{figure}

The finite time step leads to a small unnatural vibration between elements that are in contact with each other. Given the non-elastic collision scheme employed, the unnatural numeric vibration would continuously remove energy from the system. To prevent this energy loss, each element is given a small shell with a depth $S$. If two elements overlap by a distance smaller than the shell depth, the repulsive force is elastic and no energy is lost to numeric vibration. The possible computational element interactions and the forces acting on them are illustrated in Fig.~\ref{fig:two_ele}.


 Silicate and iron elements are allowed to have different repulsive strength multipliers, $K_{\rm i}$, different repulsion reduction multipliers, $\alpha_{\rm i}$, and different shell depths, $S_{\rm i}$. The minimum allowed separation distance between two elements $\epsilon$ prevents a singularity in the gravitational force and serves as a check on the simulation parameters: separation distances smaller than $\epsilon$ indicate that the repulsive strength $K_{\rm i}$ should be increased or the simulation time step should be decreased.  We conduct two different simulations using the {\sc cataclysm} code, CAT1 and CAT2. The parameters in CAT1 were set to match the results of impacts onto a fast-spinning proto-Earth proposed by \cite{Cuk2012} which resulted in angular momentum and mass ratio of the Earth-Moon system similar to the current measured values \citep{Wyatt2018}. The parameters in CAT2 were selected to match the results presented for the angular momentum presented in \cite{Canup2004}.  The values are sumarized in Table~\ref{table:Wyatt3}.  The relation between these parameters and the thermodynamics included in SPH codes is presented in the appendix along with a video of the {\sc cataclysm} results for run119.  
  

\begin{table}
\caption{Parameter Set for CATACLYSM\label{table:Wyatt3}}
\centering
\begin{tabular}{|l|l|l|}
\hline
 Parameter  & Description  & CAT1/CAT2\\
\hline
$\rho_{\rm Si}$ & Density Silicate $(\rm g/cm^{3})$ & 3.32\\
$\rho_{\rm Fe}$ & Density Iron $(\rm g/cm^{3})$ & 7.78\\
K$_{\rm Si}$ &  Repulsive strength silicate (GPa) & 150 / 290\\
K$_{\rm Fe}$ &  Repulsive strength iron (GPa) & 150 / 582\\
$\alpha_{\rm Si}$ & Percent reduction of $K_{\rm Si}$ & 0.01 / 0.01\\
$\alpha_{\rm Fe}$ & Percent reduction of $K_{\rm Fe}$ & 0.01 / 0.02\\
S$_{\rm Si}$ & Silicate shell depth (D) & 0.002 / 0.001\\
S$_{\rm Fe}$ & Iron shell depth (D) & 0.02 / 0.002\\
\hline
\end{tabular}
\par    
{$^a$ Note: Iron and silicate elements have a common diameter $D$ which is a function of the  number of elements and the total mass and composition of the system.}
\end{table}

\subsection{Analysis}
We use the same methods to analyze the outcome of the collision across all of the numerical methods used.  To compare our results with those of earlier SPH and CTH studies we used the same initial conditions as \cite[e.g.,][]{Canup2004,CanupBarr2013}, and followed the analysis method proposed by \cite{Canup2001}. Given an initial guess for the radius of the post-impact Earth, all of the material in the domain is assigned to be ``inside'' or ``outside'' of the planet based on its location.  The location and velocity of each element/particle outside the planet are used to determine the radius of its equivalent circular orbit around the protoearth, $r_{\rm circ}=(h_{\rm z}^2/GM_{\rm pl})$, where $G$ is the gravitational constant, $M_{\rm pl}$ is the mass of the planet, $h_{\rm z}$ is the angular momentum per unit mass ($h_z=xv_{\rm y}-yv_{\rm x}$, where $x$, $y$, $v_{\rm x}$, and $v_{\rm y}$ are the position of a particle and its velocity relative to the center of mass of the debris cloud.  If $r_{\rm circ}$ is less than the radius of the Earth, the particle is re-classified as ``inside'' the Earth.  This procedure is repeated iteratively until the mass of the Earth converges.  When the final Earth mass is calculated, the orbital elements of particles in the disc (i.e., those ``outside'' the planet) are computed one final time, to determine how much material has become gravitationally unbound from the system, $M_{\rm esc}$.  For run119, where $v_{\rm imp} \approx v_{\rm esc,sys}$, $M_{\rm esc}$ is typically quite small.

$N$-body gravitational simulations of the sweep-up of debris in the disc in the days, weeks, and months after the Moon-forming impact show that the mass of the final moon formed is related to the disc mass $M_{\rm disc}$  by
\begin{equation}
M_{\rm moon}\approx \frac{1.9 J_{\rm disc}}{\sqrt{GM_{\oplus}a_{\rm R}}} - 1.15 M_{\rm disc}-1.9 M_{\rm esc} \label{eq::moonmass}
\end{equation}
\citep{Ida1997}, where the scaled angular momentum 
\begin{equation}
J_{\rm disc}=L_{\rm D}/M_{\rm pl}(GM_{\oplus}a_{\rm R})^{1/2}.
\label{eq::J}
\end{equation} 
is calculated from the angular momentum of the disc $L_{\rm D}$, and $a_R\sim 2.9 R_{\oplus}$ is the Roche radius, calculated by 
\begin{equation}
     a_{\rm R}=2.456 R_{\rm pl} (\rho_{\rm disc}/\rho_{\rm pl})^{1/3},
\end{equation}
where the density of the disc material $\rho_{\rm disc}=3.3$ g/cm$^3$ is equal to the density of uncompressed silicate, and the density of the planet is equal to the density of the Earth, $\rho_{\rm pl}=5.5$ g/cm$^3$.

All of the numerical methods used in this paper permit us to track the mixing of the target with the impactor during the impact and disc evolution.  In each code, silicate originating from the protoearth and impactor are considered to be different materials, but with the same equation of state.  The same is true for iron.  In addition to determining the mass of the disc ($M_{\rm disc}$), we also determine the mass fraction of iron in the disc,
\begin{equation}
m_{\rm fe,disc} = \frac{M_{\rm fe,tar,disc}+M_{\rm fe,imp,disc}}{M_{\rm disc}},
\end{equation}
where $M_{f\rm e,tar,disc}$ is the mass of iron in the disc that originates from inside the target, and $M_{\rm fe,imp,disc}$ is the mass of iron originating from inside the impactor.  The mass fraction of terrestrial material in the disc is determined by
\begin{equation}
m_{\rm tar, disc}=\frac{M_{\rm tar,fe, disc}+M_{\rm tar,si,disc}}{M_{\rm disc}},
\end{equation}
where $M_{\rm tar,fe,disc}$ is the mass of target iron in the disc, and $M_{\rm tar,si,disc}$ is the mass of target silicate in the disc.


\section{Results}
\label{sec::results}
Any simulation of the Moon-forming impact shows that evolution proceeds in three phases: (1) the first five hours, in which the objects collide, mix, and material is lofted into orbit, (2) hours five through ten, in which orbiting clumps of material re-accrete (or not) onto the protoearth, and (3) hours ten and beyond, in which the disc evolves due to gravitational forces (and in a shock physics code, the code's innate artificial viscosity \citep{CanupBarr2013}). In this section, we compare the disc properties and lunar composition across the different computational methods.

\begin{figure}
\centering
	\includegraphics[width=0.99\columnwidth]{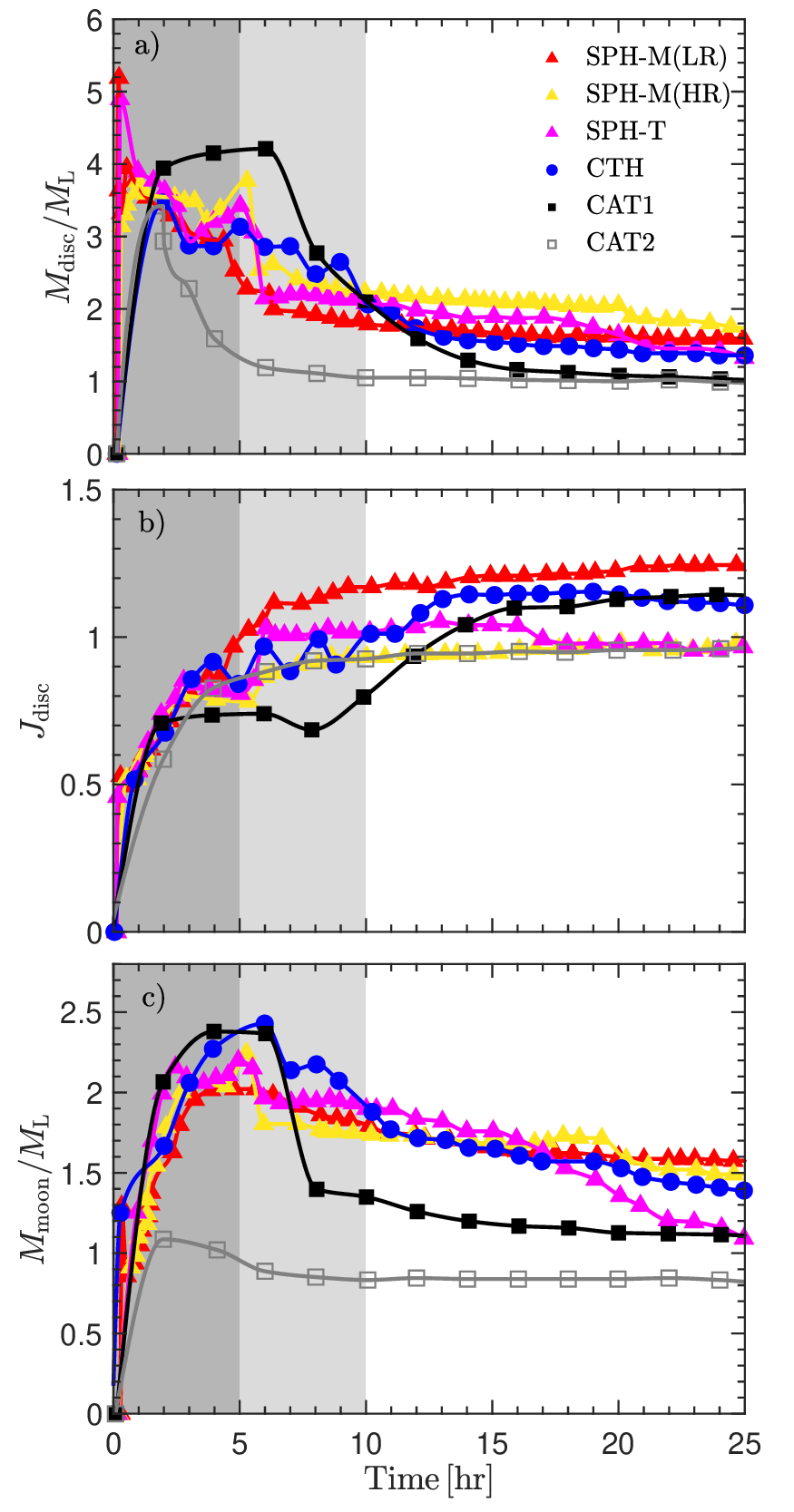}
    \caption{Comparison of disc masses (top panel), normalized disc angular momentum (middle panel), and predicted moon masses (bottom panel) for the conditions of run119 simulated using different numerical methods. The  triangles correspond to the SPH simulations which includes SPH with MANEOS based on data from \citet{CanupBarr2013} for low resolution (red, $10^4$ particles, SPH-M(LR)) and high resolution (yellow, $10^5$ particles, SPH-M(HR)), and with a Tillotson equation of state for basalt and iron (pink, SPH-T). The blue points represent CTH with MANEOS. The squares indicate the gravity plus quadratic repulsive simulations, CAT1 (black) and CAT2 (gray).  The lines fit to the data points serve to guide the eye. The shaded regions indicate the three phases during the Moon-forming impact: the first five hours in which the objects collide, mix, and material is lofted into orbit; hours five to ten, when the orbiting clumps of material re-accrete onto the protoearth; and hours ten and beyond, in which the disc around the protoearth evolves. Across all numerical methods and resolutions, the mass of the orbiting disc and disc angular momentum are broadly consistent.}
   \label{fig:timelogs}
\end{figure}

\begin{figure}
\centering
\includegraphics[width=1\columnwidth]{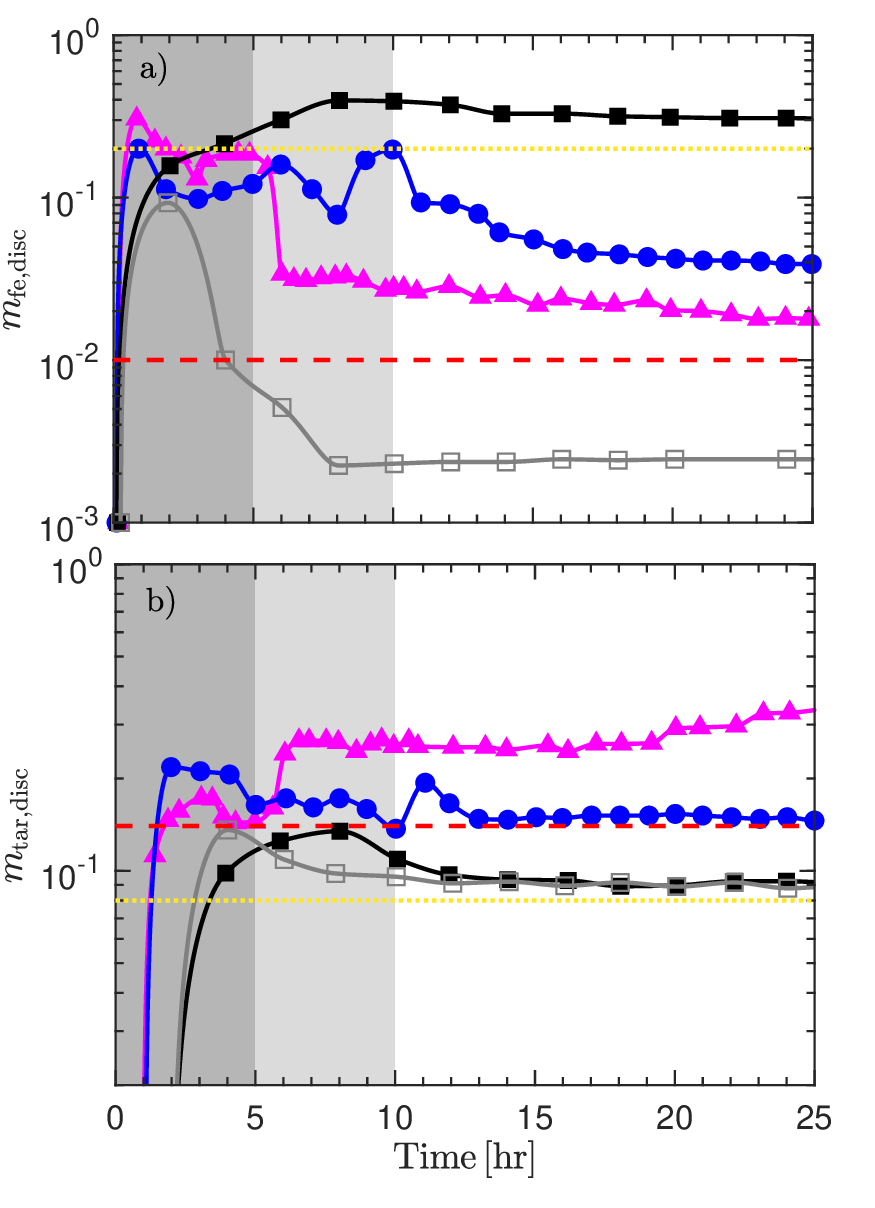}
\caption{Two measures of lunar composition: (top) mass fraction of iron in the disc (and final Moon), and (bottom) mass fraction of the disc originating in the protoearth.  Legend is the same as Fig.~\ref{fig:timelogs},  but final values for SPH with MANEOS with $N=10^4$ particles are indicated with dashed lines, and $N=10^5$ particles with dotted lines.  The mass fraction of target (Earth) material in the disc also varies significantly across all the compared simulations.  Spheral, CTH, and high-resolution SPH simulations yield the most Earth-rich discs, with the mass fraction of Earth material in the disc approaching 50 per cent for the low-resolution Spheral case.}
\label{fig:chemistry}
\end{figure}

\begin{figure*}
\centering
\includegraphics[width=2\columnwidth]{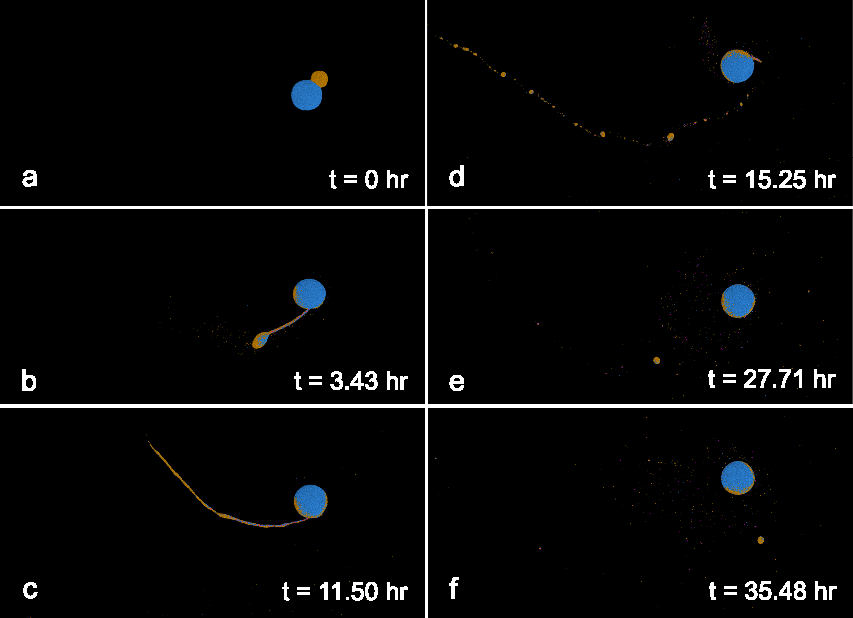}
\caption{Illustration of moon formation from the {\sc cataclsym} code, showing six different times during the impact process. (a) Initial condition for the impact, with the impactor (yellow) and the target (blue). Moon-lets are eventually formed as seen in (e) and (f).  }
\label{fig:moon}
\end{figure*}

\subsection{Disc properties}

 The top panel in Fig.~\ref{fig:timelogs} illustrates the evolution of the disc mass (scaled by the lunar mass $M_{\rm L}=7.35 \times 10^{25}\, \rm g$) as a function of time for Spheral using  MANEOS with two different resolutions ($N=10^4$ and $N=10^5$ particles), Spheral using the Tillotson equation of state,  CTH (with MANEOS), and two simulations from the gravitational code {\sc cataclysm} (CAT1 and CAT2)  for the first 25 hours after the Moon-forming impact.  The shaded regions indicate the three phases during the Moon-forming impact. Within the first five hours, material begins to form the disc proceeding from the collision of the two bodies. From hours five to ten, material is re-accreted onto the protoearth which decreases the disc mass. Beyond roughly ten hours, the disc mass is in a quasi-steady state. The differences in the evolution of disc mass in the early hours of the impact are controlled by the extent of clumping in the disc and the rate at which clumps in the disc re-impact and become absorbed onto the Earth. The decay in disc mass in all of the hydrocode simulations is controlled by the artificial viscosity inherent in each of the numerical methods; the rate of decay of disc mass is directly proportional to the strength of the viscosity parameter \citep{CanupBarr2013}. The {\sc cataclysm} code, with no artificial viscosity, throws material into orbit. When a particle impacts the Earth in {\sc cataclysm}, it merges with the Earth.  For times $t\gtrsim 10$ hours post-impact, the disc is sparse enough that there are few particle-particle collisions, and thus, the mass of the disc remains relatively constant over time. At $25$ hours post-impact, we find that all of the methods produce broadly similar disc masses.

The middle panel in Fig.~\ref{fig:timelogs} shows the scalar angular momentum of the disc (calculated from Eq.~\ref{eq::J}) as a function of time. As the disc forms from the collision, the angular momentum increases rapidly within the first five hours.  $J_{\rm disc}$ continues to increase as a function of time even as the disc mass decreases, but at different rates in each model due to the differing artificial viscosity treatments in each code. Beyond ten hours, the disc is in a quasi-steady state where the angular momentum is nearly constant. Here, we find significant differences in the final disc angular momenta, with the low-resolution SPH-M simulation yielding the highest $J_{\rm disc}$ value, while the high-resolution SPH-M, SPH-T, and CAT2 yield the smallest.

The bottom panel in Fig.~\ref{fig:timelogs} shows the estimated moon mass calculated from the disc mass using Eq.~\ref{eq::moonmass}.  We find that the predicted lunar masses vary significantly, with SPH (MANEOS) and CTH yielding the most massive moons ($\sim 1.5$ lunar masses), and CATACLYSM and SPH (Tillotson EOS) simulations yielding lunar masses $\sim M_{\rm L}$.  The use of Eq.~\ref{eq::moonmass} allows the moon mass to be estimated from the disc mass  without the simulated run times necessary to form a moon.  However, one of the strengths of the CATACLYSM code is the ability to form a lunar body without much computational cost (we discuss this further in Section 4).



\subsection{Lunar Composition}

Despite the similarities in the mass of the disc and the predicted lunar masses, we find that the moons created in each of the simulations have vastly different compositions. In all simulations of run119, the majority of iron in the impactor merges with the core of the protoearth, and very little remains in orbit, resulting in an iron-poor disc. The lunar crust is relatively iron-rich with $3-8$ percent iron mass by weight \citep{Lucey1995}, though the total fraction of iron is less than $\sim 10$ percent due to the small iron core which comprises $<2\%$ of the total lunar mass \citep{Briaud2023}.  The top panel of Figure \ref{fig:chemistry} illustrates how the mass fraction of iron in the disc varies as a function of time for SPH-T, CTH, and {\sc cataclysm}. The final values of $m_{\rm fe,disc}$ for SPH-M with $N=10^4$ and $N=10^5$ particles are indicated by the dotted and dashed lines, respectively. CTH and the SPH-M(HR) runs yield discs with an iron content close to the present lunar value, $4$ and $20$ per cent, respectively.  SPH-T  and SPH-M(LR)  yield discs that are relatively iron-poor at $2$ and $1$ per cent, respectively.


The bottom panel of Figure \ref{fig:chemistry} shows how the mass fraction of terrestrial material in the disc as a function of time varies as a function of time.  The proportions of terrestrial material in the disc are similar for CTH, SPH-M, CAT1 and CAT2, within $\sim10$\%.  SPH-T yields $>30$ percent terrestrial material in the disk, indicated that a large fraction of the material in the protoearth is lofted into orbit during the impact.



\section{Discussion and Summary}
\label{sec::summary}







 The Moon-forming impact is one of the most fundamental problems in planetary science, and one of the most-often simulated. To date, simulations of the event have been performed with perhaps a dozen different numerical methods. It is intriguing to note that although the hydrodynamic methods and the method using gravity plus inelastic spheres differ in several respects, they give similar results for the system's angular momentum and mass of the moon, which is inferred from the mass of the disc. Multiple studies, including our analysis here, show that for identical initial conditions, the mass of the impact-generated disc varies by less than $10$ per cent when the different types of simulations are performed with similar numerical resolution.  However, we find that the iron content of the disc, and the amount of terrestrial material present in the disc, which control the mean density of the final Moon and the refractory isotopes, can differ significantly. 
 
 This is consistent with other studies which found that the equation of state has a significant influence on the composition of the disc material \citep{Canup2004,Hosono2019}.   \citet{Hosono2019} proposed that the proto-Earth was covered with a magma ocean. Molten silicates are much stiffer and experience more heating under shock compression. Using a hard-sphere equation of state appropriate for modeling molten silicate produced in a larger fraction of the target material in the disc. resulting in a model which more closely matched the compositional similarities between the Moon and Earth.   

 The data also suggests that the model using gravity plus inelastic spheres could be adjusted to yield values for the mass fraction of iron and the mass fraction of the disc originating from target material which are similar to those predicted by the hydrodynamic codes.  Since the simplified code with gravity and inelastic spheres can easily be adjusted to match the orbital dynamics of the bodies resulting from a collision, we believe it could be used as a parameter space search tool. Performing an adequate search through all the possible parameters including size, composition, position, velocity, spin angle, spin velocity, etc. for both target and impactor can be a daunting task. The simplified code could be incorporated into a hyperspace tuning algorithm to find promising impact scenarios that could then be studied in more detail with hydrodynamic codes. 

 One advantage of using the modeling approach of gravity plus inelastic spheres is the reduced computational power needed to perform simulations. An example of moon formation from the {\sc cataclysm} code is shown in Fig.~\ref{fig:moon}. Panel 1 shows the initial impact of the simulation, where the target material is blue, and the impactor is yellow. Panel 2 shows a time shortly after impact when a bridge of material connects the impactor to the target. In panel 3, the majority of the impactor has accreted onto the target, leaving a tail of material. The material within the tail begins to coalesce into bound clumps in panel 4. Eventually, these clumps will merge into a Moon or accrete onto the target. In panels 5 and 6, the surviving clumps coalesce further to form moonlets.  The conditions modeled here result in the formation of two moonlets (second body not visible in Fig.~\ref{fig:moon}). However, adjusting the initial conditions can lead to the formation of a single moon, as described in \citet{Wyatt2018}.

The importance of the simplified model resides in its capacity to illustrate that the resultant system parameters, including spin, angular momentum, and mass of the bodies, are predominantly governed by gravitational forces.  The resultant system parameters are modified by altering the spring constant, a factor tied to material properties. Note that the choice of spring constant has relatively little effect on the mass of resultant disc \ref{fig:timelogs}a or the mass of the target which ends up in the disc \ref{fig:chemistry}b. The relative compositions of the resultant bodies will depend on the equation of state which captures the material changes occurring during collisions. The spring model represents a highly simplified model of a given Equation of State (EoS) which can capture the energy loss as material is heated during collisions (see Appendix~\ref{appendix::a}).  It is important to note that a perfect match in this regard is not achievable, but the energy loss modeled in the simplified models do reasonably approximate final temperatures obtained from models with more sophisticated equations of state.

An illustration of how the spring constant can influence the final compositions following the encounter is evident in Fig. \ref{fig:chemistry}a. The gravity-only models, CAT1 and CAT2, yield the highest and lowest values for the proportion of iron in the resulting disc. The outcomes are contingent upon the selected values for the repulsive constant $K_{\rm Fe}$, reduction of energy $\alpha_{\rm Fe}$, and shell depth $S_{\rm Fe}$. In the case of CAT1, with lower values of $K_{\rm Fe}$ and $\alpha_{\rm Fe}$, the disc exhibits an excessive iron content, surpassing 30 per cent. In contrast, CAT2 represented very stiff spheres with a value of $K_{\rm Fe}$ 3.88 larger than in CAT1 and a much thinner shell depth.  The resultant disc iron fraction in this model is approximately $2\times 10^{-3}$, which is 40 times less than that present in the Moon. This suggests that by appropriately adjusting the parameters $K_{\rm Fe}$ and $S_{\rm Fe}$, the {\sc cataclysm} model could more accurately reflect the underlying physics of the EoS and yield a more appropriate value for the proportion of iron.

In conclusion, the formation and characteristics of solid planetary bodies, including their sizes, compositions, and angular momenta, hinge upon the outcomes of collisions between planetary embryos. While the prevailing numerical method involves combining gravity and hydrodynamics solvers to account for complex phenomena like pressure gradients, shock waves, and gravitational torques, this study marks directly compares this approach with a simplified model relying solely on gravity and a quadratic repulsive force. Focusing on the highly-studied case of the formation of Earth's Moon, we've shown that many fundamental aspects of planetary embryo collisions, such as the mass and orbit of satellites and the degree of physical mixing, remain consistent irrespective of the inclusion of hydrodynamic effects or the specific equation of state. Nonetheless, to fully comprehend thermal and chemical effects and determine the timescales involved, full hydrodynamic calculations are essential. This simplified approach offers a valuable tool for rapidly exploring diverse initial conditions, helping pinpoint cases warranting more detailed investigation.

\section*{Acknowledgements}
We express our appreciation to Amy C. Barr-Milner for the invaluable contributions they have made to our research project. Amy C. Barr-Milner initiated the original concept of model comparison and generously provided crucial data for both the Spheral and CTH codes. Their perspectives and data have played a pivotal role in guiding the focus of our study and elevating the caliber of our results. JLS acknowledges funding from the ASIAA Distinguished Postdoctoral Fellowship. BMW and LSM thank NVIDIA for hardware donations and BMW thanks Tarleton University's HPC lab.

 \section*{Data Availability}
 The github repository of the {\sc cataclysm} code can be downloaded at \url{https://github.com/TSUParticleModelingGroup/Giant-Impact.git}. Link to video of {\sc cataclysm} running Canup's 119 initial conditions. 
 \url{https://www.youtube.com/watch?v=isMEeTTEY8E}.

\section*{Credit author statement}
{\bf Jeremy L. Smallwood:} Visualization, Writing – Review \& Editing. {\bf Jeff S. Lee:} Formal analysis.
{\bf Lorin S. Matthews:} Formal Analysis, Writing – Review \& Editing. {\bf Bryant M. Wyatt:} Conceptualization, Methodology, Software.



\bibliographystyle{mnras}
\bibliography{ref} 




\appendix

\section{Calculation of Inelastic sphere parameters}
\label{appendix::a}
There are three constraints which a successful model of moon formation must satisfy: the resultant angular momentum of the system must be close to the current value, the collision must produce a disk with sufficient mass to form the moon, and the chemical composition of the material forming the moon must agree with present-day observations.  The initial conditions determining the impact kinetics control the angular momentum and disk mass to a large degree.  The choice of equation of state has been shown to have a significant influence on the composition of the disk formed by a giant impact \citep{Canup2008, Hosono2019}. 

The constants in the {\sc cataclysm} code are tuned so that results match the resultant orbital parameters and disk mass of other impact models. Here we show that these constants are also consistent with the thermodynamics modeled by an equation of state. In the CATACLYSM model, the kinetic energy of two colliding spheres is transformed into potential energy of a compressed spring in an inelastic collision and is subsequently lost as heat.  The choice of the spring constants $K_{\rm Si}$  and $K_{\rm Fe}$ as well as the reduction factors for the spring constants $\alpha_{\rm Si}$  and $\alpha_{\rm Fe}$ reflect the simplified underlying physics of a particular equation of state. 

\begin{figure}
\centering
\includegraphics[width=\columnwidth]{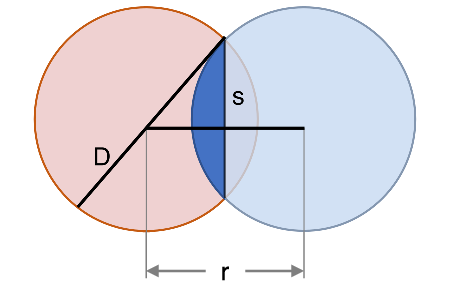}
\caption{Collision between two spheres of diameter $D$. The distance between the centers of the two sphere is $r$. The contact area between the two spheres is a circle of radius $s$. }
\label{fig:sphere_collision}
\end{figure}

A simplified model of two colliding spheres is depicted in Fig.~\ref{fig:sphere_collision}.  The pressure resulting from the impact of two spheres is given by $P=F/A$ where $F$ is the force between the two spheres and $A$ is the area of contact. 

In the spring model, the force is proportional to the area of the spheres in contact, $F=kA$. In this case, the pressure during the collision is constant and $P=k$.  The contact area is $A=\pi s^2$, where $s$ is the radius of the circular contact area, which can be written as a function of $r$, the separation between the centers of the two spheres, $A = \frac{\pi}{4}(D^2 - r^2)$, where $D$ is the diameter of each sphere (all spheres have the same diameter in the {\sc cataclysm} model).

To simplify calculations in the code,  the factor of $\pi/4$ is included in the effective spring constant, such that
\begin{equation}
F = (K_1 + K_2)(D^2 - r^2),
\end{equation}
where $K_1$ and $K_2$ are the effective spring constants of each of the spheres. The potential energy of the compressed spring system is found by integrating the force, resulting in 
\begin{equation}
    U = - (K_1 + K_2) \int_D^r (D^2 - r'^2) dr'.
\end{equation}
The minimum separation distance between the centers of the two spheres $r_{\rm min}$ is found by setting the kinetic energy at the point of impact equal to the potential energy at maximum compression
\begin{equation}
    2(\frac{1}{2}mv^2) = (K_1 + K_2) \bigg( \frac{2}{3}D^3 - D^2 r_{\rm min} + \frac{1}{3}r_{\rm min}^3\bigg),
\end{equation}
where we have assumed two spheres of equal mass $m$ are approaching each other with equal and opposite velocities $v$ in the center of mass system. $K_1$ and $K_2$ must be great enough that the spheres do not pass through each other so that $r_{min}> D/2$. 

As a test case, we consider a collision between two silicate spheres each with a mass of $3.77\times10^{19}$ kg and diameters  $D = 279$ km.  Assuming that the two sphere approach each other with a closure velocity of $9.4$ km/s (the velocity between the proto-earth and impactor just before collision), each sphere has a velocity of $4.7$ km/s.  Upon collision, the minimum distance between the centers of the spheres is $169$ km for the conditions in CAT1 and $202$ km for the condition in CAT2.

The energy of the collision is transformed into heat. In the {\sc cataclysm} model, this energy loss is represented by the inelastic collision between the two spheres, with the spring constant reduced when the spheres are receding are receding from each other,
\begin{equation}
K_i^{'} = \alpha_i K_i,  
\end{equation}
where $\alpha_i$ is the reduction factor.  The energy {\it available} to heat the spheres is the difference in the total kinetic energy of both spheres before and after the collision  
\begin{equation}
\Delta E_k = (1-\alpha_i )(mv_i^2).
\end{equation}
The energy that actually does heat the spheres is $E_h = \beta \Delta E_k$, where $\beta$ is the Taylor-Quinney coefficient. Appropriate published values of $\beta$ range from 0.86-0.90.  In a collision of two spheres, each sphere should receive half of this energy to heat the material.

 Using the example of two silicate spheres colliding with a velocity just before the point of impact of 4.7 km/s,the energy available to heat two spheres is 
\begin{equation}
E_h = \beta \Delta E_k=\beta (1-\alpha_i) mv^2.
\end{equation}

The spheres are taken to be thermodynamically homogeneous elements with the same temperature throughout. The pre-collision state of a sphere may be solid or liquid (in the case of a molten magma ocean \citep{Hosono2019}.  For solid spheres, we assume that the Hadean (or Archean) mantle potential temperature $T_p$ is the initial temperature of the spheres when the bodies are broken apart. Estimates vary, with $T_p$ $\sim 1770-1870~K$ \citep{hawkesworth2016tectonics,Tappe2018,Weller2019}.  \citet{Hosono2019} assumed a molten ocean covering the surface of the target with an initial temperature of 2000 K. Given an initial temperature of 1770 - 2000 K, $\alpha_Si= 0.01$, and $\beta = 0.86 - 0.90$ the impact energy being converted into heat results in a final temperature of  8000 - 9000 K.  These are consistent with the temperatures reported in previous studies \citep{Canup2008, CanupBarr2013, Hosono2019}.

\bsp	
\label{lastpage}
\end{document}